\documentstyle[psfig,conf_iap,]{article}
\begin{document}
\heading{%
%
Generalized Chaplygin gas as a scheme for Unification of 
Dark Energy and Dark Matter
%
} 
\par\medskip\noindent
\author{%
M.C. Bento$^{1,2}$, O. Bertolami$^{1,a}$, A.A. Sen$^{3}$ 
}
\address{%
Departamento de F\'\i sica, Instituto Superior T\'ecnico \\
Av. Rovisco Pais 1, 1049-001 Lisboa, Portugal
}
\address{Centro de F\'{\i}sica das 
Interac\c c\~oes Fundamentais, Instituto Superior T\'ecnico}
\address{Centro Multidisciplinar de Astrof\'{\i}sica, 
Instituto Superior T\'ecnico}

\noindent
$^{a}$ Speaker

\begin{abstract}
We study the cosmological scenario arising from the dynamics of a 
generalized Chaplygin gas. The equation of state of the 
system is given in terms of the energy density, $\rho$, and 
pressure, $p$, by the relationship $p = - A/\rho^{\alpha}$, where $A$ 
is a positive constant and $0 < \alpha \le 1$. The conditions 
under which homogeneity arises are discussed and it is shown 
that this equation of state 
describes a universe evolving from a phase dominated by 
non-relativistic matter to a phase dominated by a cosmological constant via an 
intermediate period where the effective equation of state is $p = \alpha \rho$.
\end{abstract}
\section{Introduction}
Recently, it has been suggested that the change of behaviour of the missing 
energy density might be controlled by the change in the equation 
of state of the background fluid instead of the form of the potential, 
avoiding in this way fine-tuning problems \cite{Kamenshchik}. In the framework of 
Friedmann-Robertson-Walker cosmology, this is achieved considering an exotic 
background fluid, the Chaplygin gas, described by the equation of state

\begin{equation}
p = - {A \over \rho^\alpha}~~,
\label{eqstate}
\end{equation}
with $\alpha=1$ and A a positive constant. Introducing this equation of state 
into the equation resulting from the covariant conservation of the energy-momentum tensor 
for an homogeneous and isotropic spacetime, leads to a density evolving
as

\begin{equation}
\rho =  \left(A + {B \over a^{3 (1 + \alpha)}}\right)^{1 \over 1 + \alpha}~~,
\label{genevolden}
\end{equation} 
where $a$ is the scale factor of the Universe and $B$ an integration 
constant. It is remarkable that this simple model smoothly
interpolates between a dust dominated phase where 
$\rho \simeq \sqrt{B} a^{-3}$ and  
a De Sitter phase where $p \simeq - \rho$, 
through an intermediate regime described, for $\alpha = 1$,
by the equation of state for  stiff matter, $p = \rho$ \cite{Kamenshchik}.
It is interesting that the Chaplygin gas admits a brane interpretation as
Eq. (\ref{eqstate}), with $\alpha = 1$, is the equation of state associated 
with the parametrization invariant Nambu-Goto $d$-brane 
action in a $(d+1, 1)$ spacetime. This action leads, in the light-cone 
parametrization, to the Galileo-invariant Chaplygin gas in a $(d, 1)$
spacetime 
and to the Poincar\'e-invariant Born-Infeld action in a $(d, 1)$ spacetime 
(see \cite{Jackiw} and references therein).  

In what follows we discuss the results of our research on the 
generalized Chaplygin gas with $0 < \alpha \le 1$ \cite{Bento}.

\section{The Model}

We consider, as first discussed in Ref. \cite{Bilic}, the Lagrangian density 
for a massive complex scalar field, $\Phi$, 
\begin{equation}
{\cal L} = g^{\mu \nu} \Phi^{*}_{, \mu} \Phi_{, \nu} - V(\vert \Phi \vert^2)~~.
\label{complexfield}
\end{equation}
This scalar field, with mass $m$, 
can be expressed 
as $\Phi = (m \phi / \sqrt{2}) \exp(- im \theta)$. 

The scale of the inhomogeneity is set assuming that 
spacetime variations of  $\phi$ correspond to scales greater than 
$m^{-1}$, thus

\begin{equation}
\phi_{, \mu} << m \phi~~.
\label{inhom}
\end{equation}
In this (Thomas-Fermi) approximation, the Lagrangian density can  
be written as 

\begin{equation}
{\cal L}_{TF} = {\phi^2 \over 2} g^{\mu \nu} \theta_{, \mu} \theta_{, \nu} 
- V(\phi^2/2)~~.
\label{Thomas-Fermi}
\end{equation} 
 
\noindent
Notice, that the field $\theta$ can be regarded as a 
velocity field provided $V'> 0$, i.e.

\begin{equation}
U^{\mu} = {g^{\mu \nu} \theta_{, \nu} \over \sqrt{V'}}~~,
\label{velocity}
\end{equation}
so that, on the mass shell, $U^{\mu} U_{\mu} = 1$. Hence, the 
energy-momentum tensor built from the Lagrangian density 
Eq. (\ref{Thomas-Fermi}) takes the form of a perfect fluid whose 
thermodynamic variables can be written as 

\begin{equation}
\rho = {\phi^2 \over 2} V' + V~~, 
\label{pardensity}
\end{equation}
 
\begin{equation}
p = {\phi^2 \over 2} V' - V~~. 
\label{parpressure}
\end{equation}

Using Eqs. (\ref{pardensity}) and (\ref{parpressure}),
together with Eq. (\ref{eqstate}), one finds a relationship between 
$\phi^2$ and $\rho$:
\begin{equation}
\phi^2(\rho) = \rho^{\alpha} 
(\rho^{1 + \alpha} - A)^{{1 - \alpha \over 1 + \alpha}}~~.
\label{phidens}
\end{equation}

Further manipulation, substituting Eqs. (\ref{pardensity}), 
(\ref{parpressure}) 
and (\ref{phidens}) into the Lagrangian density (\ref{Thomas-Fermi}),
shows that it is possible to establish a brane connection to this model, 
as the resulting Lagrangian density has 
the form of a {\it generalized} Born-Infeld theory:

\begin{equation}
{\cal L}_{GBI} = - A^{1 \over 1 + \alpha} 
\left[1 - (g^{\mu \nu} \theta_{, \mu} 
\theta_{, \nu})^{1 + \alpha \over 2\alpha}\right]^{\alpha \over 1 + \alpha}~~.
\label{GenBorn-Infeld}
\end{equation} 

The potential arising from this model can be written as

\begin{equation}
V = {\rho^{1 + \alpha} + A \over 2 \rho^{\alpha}} = 
{1 \over 2} \left(\Psi^{2/\alpha} + {A \over \Psi^2}\right)~~,
\label{pot}
\end{equation}
where $\Psi \equiv B^{-(1 - \alpha/1 + \alpha)} a^{3(1 - \alpha)} \phi^2$, 
which reduces to the duality invariant, $\phi^2 \rightarrow A/\phi^2$, 
and scale-factor independent potential for the 
Chaplygin gas.

The effective equation of state in the intermediate phase between the 
dust dominated phase and the De Sitter phase is obtained expanding 
Eq. (\ref{genevolden}) in subleading order:

\begin{equation}
\rho \simeq A^{1 \over 1 + \alpha} + \left({1 \over 1 + \alpha}\right) 
{B \over A^{\alpha \over 1 + \alpha}} a^{-3(1 + \alpha)}~~,
\label{effecden}
\end{equation} 

\begin{equation}
p \simeq - A^{1 \over 1 + \alpha} + \left({\alpha \over 1 + \alpha}\right) 
{B \over A^{\alpha \over 1 + \alpha}} a^{-3(1 + \alpha)}~~,
\label{effecpres}
\end{equation} 
which corresponds to a mixture of  vacuum energy density  
$A^{1 \over 1 + \alpha}$ 
and matter described by the ``soft'' equation of state:

\begin{equation}
p = \alpha \rho~~.
\label{effeceqstate}
\end{equation} 

In broad terms, the comparison between the cosmological setting we propose 
and the one emerging from the Chaplygin gas, discussed in Refs. 
\cite{Kamenshchik,Bilic}, 
is exhibited in Figure 1. 

\begin{figure}
\centerline{\vbox{
\psfig{figure=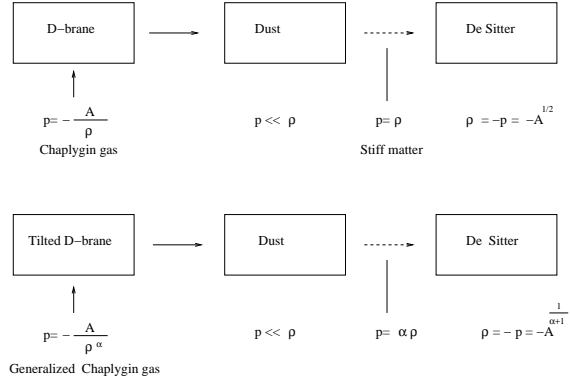,height=5cm}
}}
\caption[]{Cosmological evolution of a universe described by 
a generalized Chaplygin gas equation of state}
\end{figure}

\begin{figure}
\centerline{\vbox{
\psfig{figure=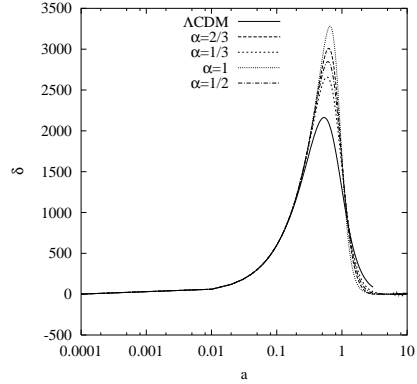,height=5cm}
}}
\caption[]{Density contrast for different values of $\alpha$, 
as compared with $\Lambda$CDM as a function of $a$ relative to 
the present ($a_0 = 1$)}
\label{fig:pert}
\end{figure}

Next, we analyse the issue of the growth of inhomogeneities in our model.
Using the Zeldovich method for the treatment of inhomogeneities, one can write,

\begin{equation}
\rho \simeq \bar \rho (1 + \delta)~~~,~~~p \simeq - 
{A \over \bar \rho^{\alpha}} (1 - \alpha \delta)~~,
\label{pert}
\end{equation} 
where $\bar \rho$ is given by Eq. (\ref{genevolden}), 
$\delta$ is the density contrast. 
Using the perturbed Friedmann equations, and the 
unperturbed Raychaudhuri equations, 
we have solved $\delta$ \cite{Bento} as a function of $a$ for different 
values of $\alpha$ which we have plotted in Figure 
\ref{fig:pert}. Hence, we verify for any $\alpha$ the claim of 
Refs. \cite{Bilic,Fabris}, for $\alpha = 1$, that 
the density contrast decays for large $a$.
Figure \ref{fig:pert} also shows the main difference in behaviour of the 
density contrast between a 
universe filled with matter with a ``soft'' or ``stiff'' equations of state 
as the former resembles more closely the $\Lambda$CDM. Research on further 
observational implications of our results is under way and will be presented 
elsewhere.
 
%
\vfill
\end{document}